\documentclass[journal=ancac3,manuscript=article,layout=twocolumn]{achemso}
\setkeys{acs}{articletitle = true} 

\UseRawInputEncoding
\usepackage[utf8]{inputenc}
\usepackage{array}
\usepackage{stfloats}
\usepackage{comment}
\usepackage{chemformula} 
\usepackage[T1]{fontenc} 
\usepackage{float}
\usepackage{seqsplit} 
\usepackage{graphicx}
\usepackage{booktabs} 
\usepackage{multirow}
\usepackage{multicol}
\usepackage{color, colortbl}
\usepackage[para,flushleft]{threeparttable}
\usepackage{ulem}
\usepackage[super]{nth}
\usepackage{amsmath}
\usepackage[font=scriptsize,labelfont=bf]{caption}
\usepackage[switch]{lineno}
\usepackage{breqn}
\usepackage{adjustbox}
\usepackage{makecell} 
\DeclareUnicodeCharacter{2009}{\,} 
\DeclareUnicodeCharacter{00C9}{\'E}
\DeclareUnicodeCharacter{00FC}{\"u}
\usepackage{chapterbib}
\mciteErrorOnUnknownfalse
\usepackage{amssymb}
\usepackage[version=4]{mhchem}
\usepackage{subcaption}
\usepackage{bbm}
\usepackage[shortlabels]{enumitem}
\usepackage{wrapfig}

\usepackage{etoolbox} 
\makeatletter
\renewcommand*{\acs@author@fnsymbol@symbol}[1]{%
  \ifcase#1 *\or 1\or 2\or 3\or 4\or 5\or 6\or 7\or 8\or 9\or 10\fi
}

\patchcmd{\acs@address@list@auxii}
  {\acs@author@fnsymbol{\acs@affil@marker@cnt}}
  {\textsuperscript{\acs@author@fnsymbol{\acs@affil@marker@cnt}}}
  {}{}

\patchcmd{\acs@address@list@auxii}
  {{\acs@author@fnsymbol{\acs@affil@marker@cnt}\@nameuse{@altaffil@\@roman\@tempcnta}\par}}
  {{\textsuperscript{\acs@author@fnsymbol{\acs@affil@marker@cnt}}\@nameuse{@altaffil@\@roman\@tempcnta}\par}}
  {}{}
\makeatother

\usepackage{titletoc} 
\author{Binay P. Nayak}
\affiliation{Department of Chemical and Biological Engineering, Iowa State University, Ames, IA, 50011, US}
\alsoaffiliation{Ames National Laboratory, Iowa State University, Ames, IA, 50011, US}

\author{Wenjie Wang}
\affiliation{Division of Materials Sciences and Engineering, Ames National Laboratory, Ames, IA, 50011, US}

\author{Prapti Kakkar}
\affiliation{Department of Chemical and Biological Engineering, Iowa State University, Ames, IA, 50011, US}
\alsoaffiliation{Ames National Laboratory, Iowa State University, Ames, IA, 50011, US}

\author{Honghu Zhang}
\affiliation{National Synchrotron Light Source II, Brookhaven National Laboratory, Upton, NY, 11973, USA}

\author{Zinnia Mallick}
\affiliation{Department of Nanoscience and Biomedical Engineering, South Dakota School of Mines and Technology, Rapid City, SD, 57701, USA}

\author{Shan Zhou}
\affiliation{Department of Nanoscience and Biomedical Engineering, South Dakota School of Mines and Technology, Rapid City, SD, 57701, USA}

\author{Dmytro Nykypanchuk}
\affiliation{Center for Functional Nanomaterials, Brookhaven National Laboratory, Upton, NY, 11973, USA}

\author{Surya K. Mallapragada}
\affiliation{Department of Chemical and Biological Engineering, Iowa State University, Ames, IA, 50011, US}
\alsoaffiliation{Ames National Laboratory, Iowa State University, Ames, IA, 50011, US}
 
\author{Alex Travesset}
\email{trvsst@ameslab.gov}
\affiliation{Ames National Laboratory, Iowa State University, Ames, IA, 50011, US}
\alsoaffiliation{Department of Physics and Astronomy, Iowa State University, Ames, IA, 50011, US}

\author{David Vaknin}
\email{vaknin@ameslab.gov}
\affiliation{Ames National Laboratory, Iowa State University, Ames, IA, 50011, US}
\alsoaffiliation{Department of Physics and Astronomy, Iowa State University, Ames, IA, 50011, US}

\title{Valence-Free Open Nanoparticle Superlattices}

\begin{document}

\vspace{-0.5 cm}

\begin{abstract}
\vspace{-0.2 cm}
A cornerstone of advanced materials design is establishing a framework for assembling nanoparticle superstructures with tailored symmetries. A longstanding challenge has been assembling diamond-like superstructures for photonic devices. Traditionally, such open superstructures require functionalized nanoparticles with directional or anisotropic interactions, reminiscent of valence bonding in a diamond. Here, we present a robust strategy for assembling valence-free nanoparticles into a broad array of cubic superstructures. By grafting nanoparticles with oppositely charged, end-functionalized water-soluble polymers of adjustable molecular weight, we gain control over electrostatic interactions and conformational constraints. This unified approach yields lattices analogous to rock salt, CsCl, zinc‑blende, diamond, and the rare simple cubic phase, with tunable lattice constants. Theoretical models and simulations elucidate the underlying interactions, providing a framework for engineering valence-free nanoparticle superlattices.

\end{abstract}

\vspace{-0.5 cm}
\section{Introduction}
\vspace{-0.2 cm}

\begin{figure*}[!hbt]
 	\centering 
 	\includegraphics[width=0.8 \linewidth]{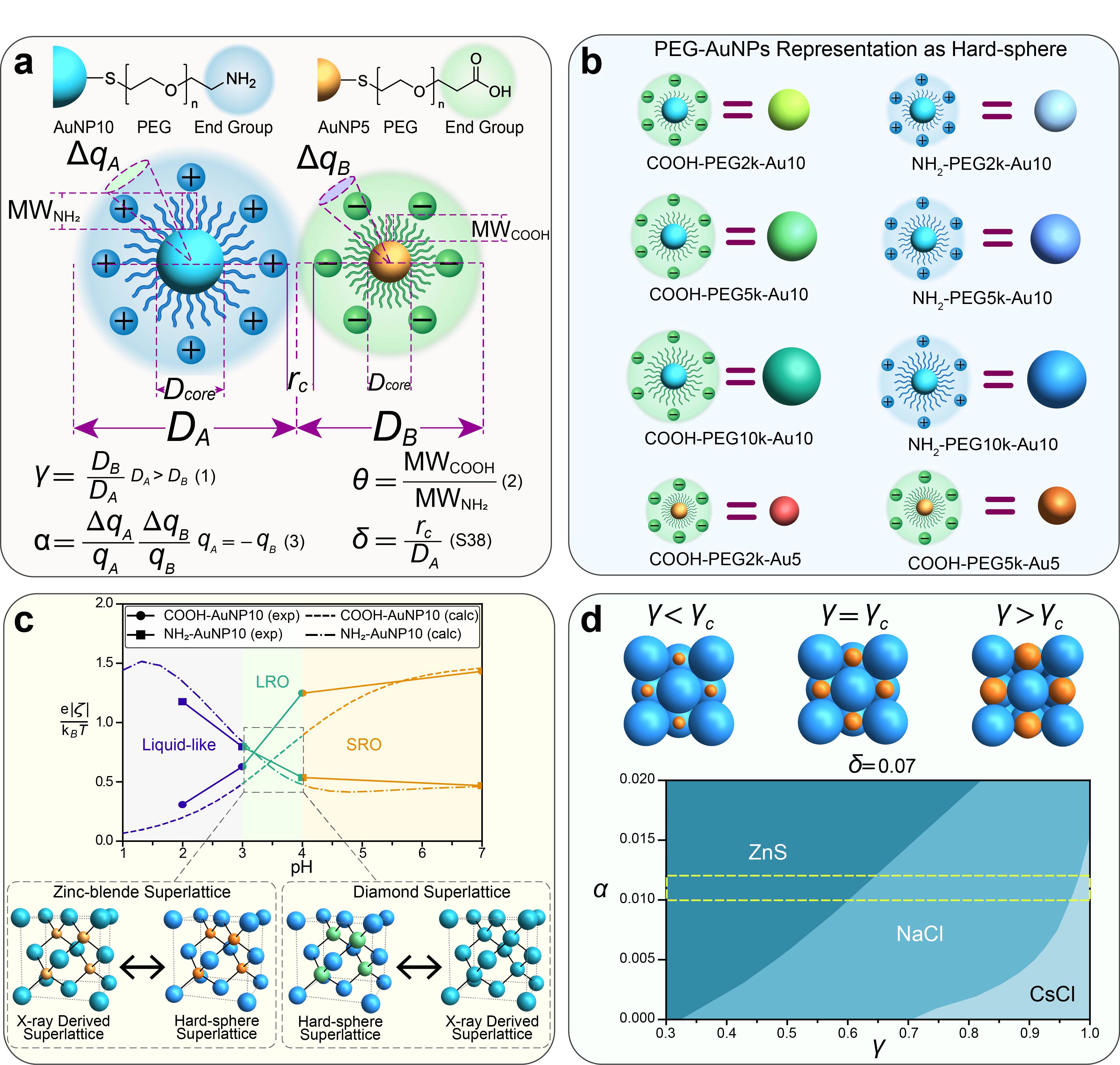}
   \caption{\textbf{Strategy to construct valence-free ionic-like NP superlattices.} (a) Depiction of gold nanoparticles (AuNPs) with core diameters ($D_{\rm core}$) of 5~nm (yellow)  and 10~nm (blue) are functionalized with poly(ethylene glycol) (PEG) terminated with either \ch{-NH2} (blue) or \ch{-COOH} (green). The surface charges on the blue ($+$) and the green ($-$) PEG corona are indicated consistently. \(D_A\) and \(D_B\) denote the actual hard sphere diameters of grafted AuNPs, with their ratio defined as \(\gamma = \frac{D_B}{D_A}\). $\Delta q_A$ and $\Delta q_B$ account for local charge correlations, and the parameter $\alpha$ measures their combined effect on stabilizing the lattice. \(\theta\) represents the molecular weight (MW) ratio of the grafted PEG chains. $\delta = \frac{r_c } {D_A}$ represents the normalized cut-off distance for nearest electrostatic interactions. (b) Legend for identifying the various NPs used to assemble the reported superlattices. The spheres on the right depict the effective hard sphere diameter after grafting AuNP with 2, 5, and 10 kDa PEG. (c) (Top) Normalized \(\zeta\)-potential as a function of pH for PEG-AuNPs, with PEG terminated by either \ch{-NH2} (squares) or \ch{-COOH} (circles). Dashed and dash-dotted lines, representing predictions from electrostatic model calculations (detailed in the SI), are also shown. Long-range ordered (LRO) superlattices form optimally at pH 3–4, while higher pH values lead to short-range order (SRO) assembly. (Bottom) Ideal zinc-blende and diamond structures derived from X-ray analysis are presented alongside the corresponding actual grafted-AuNP assemblies depicted in b. In the zinc-blende structure, distinct colors highlight the two different particle types, whereas in the diamond structure, the spheres are shown in a uniform color, reflecting that the X-ray pattern arises predominantly from identical AuNP cores. Henceforth, structures with uniform core sizes (e.g., diamond, simple cubic, and body-centered cubic) are depicted with their actual grafted PEG using the symbols from b.
   (d) (Top) (110) plane view of a rock-salt lattice for different \(\gamma\) values, illustrating ideal hard sphere contacts at \(\gamma = \gamma_c\). (Bottom) A proposed phase diagram based on geometrical constraints and electrostatic interactions that guide the assembly of various superlattices. The dashed lines indicate the region that aligns with our experimental observations.
    }
\vspace{-0.3 cm}
\label{fig:schematics} 
 \end{figure*}
 
Assembling colloidal nanoparticles (NPs) into cubic superstructures has long been a central challenge in nanoscience\cite{macfarlane2011nanoparticle,macfarlane2013topotactic,shevchenko2006structural,kostiainen2013electrostatic,nykypanchuk2008dna,Bassani2024,bian2021electrostatic,Boles2016,coropceanu2022self}. For example, the fabrication of diamond-like superstructures has traditionally relied on patchy or anisotropic NPs, with directional interaction sites\cite{Liu2016a,he2020colloidal}. The valence-like bonding mimics the directional bonding observed in carbon-based diamond structures. Various methods have been employed to achieve this goal, including the arrangement of NPs within tetrahedral DNA origami cages~\cite{Liu2016a} and the assembly of micron-sized tetrahedral clusters linked through DNA~\cite{he2020colloidal}. DNA origami tetrapods have also been used to create diamond lattices with photonic band gaps~\cite{Posnjack2024}. Low-coordination cubic lattices have also been realized through DNA-programmed NPs.\cite{mao2023regulating} Only one report of a diamond superlattice, assembled without valence, exhibited broad X-ray diffraction peaks, signaling poor structural quality~\cite{Kalsin2006}. Given these challenges, we present a general strategy for assembling a diverse array of NP superlattices, including diamond, inspired by binary ionic compounds, achieved without relying on valence.

Binary ionic compounds exhibit a rich diversity of ordered structures dictated by ion size.\cite{aguado1997structure,smart2012solid} For example, similar-size ions such as Cs$^+$ and Cl$^-$ tend to adopt simple cubic or pseudo-body-centered cubic (BCC)-like arrangements. For moderate size asymmetry, as seen with Na$^+$/K$^+$ and Cl$^-$, the prevailing configuration is a rock salt or face-centered cubic (FCC)-like structure. Divalent compounds with higher size asymmetry, from oxides like \ch{CaO} and \ch{BaO} to chalcogenides such as \ch{ZnS}, follow a similar trend.\cite{hull1921crystal} \ch{ZnS} can crystallize in a zinc-blende (FCC) or wurtzite (hexagonal) lattice, depending on external conditions.\cite{desgreniers2000pressure} Hypothetically, if the rock salt, cesium chloride, or zinc-blende structures were made with identical elements, they would correspond to simple cubic, BCC, and diamond lattices, respectively.

Building on these structural insights, we now explore how similar principles apply to the assembly of spherical NPs. We functionalize NPs with poly(ethylene glycol) (PEG) chains terminated with carboxyl (\ch{-COOH}) or amine (\ch{-NH2}) end-groups, enabling control over their surface charge through adjustment of pH. This strategy is broadly applicable, as both the NP type and the terminal groups can be generalized to any system employing ligands with oppositely charged ends. Moreover, by varying the molecular weight of PEG, we tune the polymer chain length and, consequently, the hydrodynamic size of the grafted NPs. The resulting superlattices consist of NP cores embedded within a uniform polymeric matrix. When the NP cores are identical, the structures can adopt simpler forms such as BCC, simple cubic, or diamond arrangements, which are analogous to elemental binary ionic salts, depending on their hydrodynamic size. The key insight is that the hydrodynamic radius, tuned via polymer chain length, sets the effective ion size, and thus the structure and lattice constants are independent of the NP core diameter. Early experiments with micron‑sized charged colloids in organic solvents yielded a range of ionic‑like superlattices but did not produce ZnS superlattices, a precursor for assembling diamond superlattices.\cite{Leunissen2005} However, theoretical work has shown that valence‑free NPs, interacting via designed spherically symmetric potentials, can assemble into open lattices such as diamond and simple cubic. \cite{Rechtsman2007,Rechtsman2006a}.

To further illustrate our assembly strategy, we model the grafted NPs as hard spheres with diameters \(D_A\) and \(D_B\) (with \(D_A > D_B\)) and uniformly opposite surface charges (\(q_A\) and \(q_B\)). As shown in Fig.~\ref{fig:schematics}a, the key parameter is the size ratio;
\begin{equation}\label{gamma}
     \gamma = \frac{D_B}{D_A} \leq 1.
\end{equation}
Experimentally, $\gamma$ is determined as the ratio of the effective modal hydrodynamic diameters ($D_{\rm H}$) of the two NP populations. This $\gamma$ parameter has also been examined in the context of size‐asymmetric binary charged colloidal suspensions (see Supplementary Information (SI) for details). \cite{lin2022superionic} 
\noindent Most ionic superstructures, including the ones discussed above-CsCl, NaCl, and ZnS, exhibit a critical gamma (\(\gamma_c\)) such that for \(\gamma > \gamma_c\) each \textit{B} NP is surrounded by \textit{A} NPs, stabilizing the lattice, while for \(\gamma \leq \gamma_c\) similar charged \textit{A} NPs come into contact, leading to structural instability (see Fig.~\ref{fig:schematics}d). Detailed calculations in the SI further reveal that the stability of the ZnS structure is significantly enhanced, achieving the highest asymmetric mixture, by tuning the repulsion between like-charged NPs, thereby facilitating the assembly of open structures such as diamond or simple cubic lattices.

Recent two-dimensional (2D) assembly experiments for the same binary systems indicate that crystallization occurs within a narrow pH range, typically between 2 and 5.\cite{nayak2023ionic} Furthermore, \(\zeta\)-potential measurements shown in Fig.~\ref{fig:schematics}c reveal a regime where NP charges overlap optimally for assembly, and theoretical work demonstrates maximum attraction when NP satisfies \(q_A = -q_B\). Even if the charges per particle are initially imbalanced, the proximity of the NPs induces a self-regulation that drives the system towards equal and opposite charges, thereby maximizing the attraction~\cite{Travesset2024a}. This approach
leads to assembling a wide range of NP superlattices by controlling the effective NP charge and hydrodynamic size.

\vspace{-0.4 cm}
\section{Results and Discussions}

\begin{figure*}[!hbt]
 	\centering 
 	\includegraphics[width=1\linewidth]{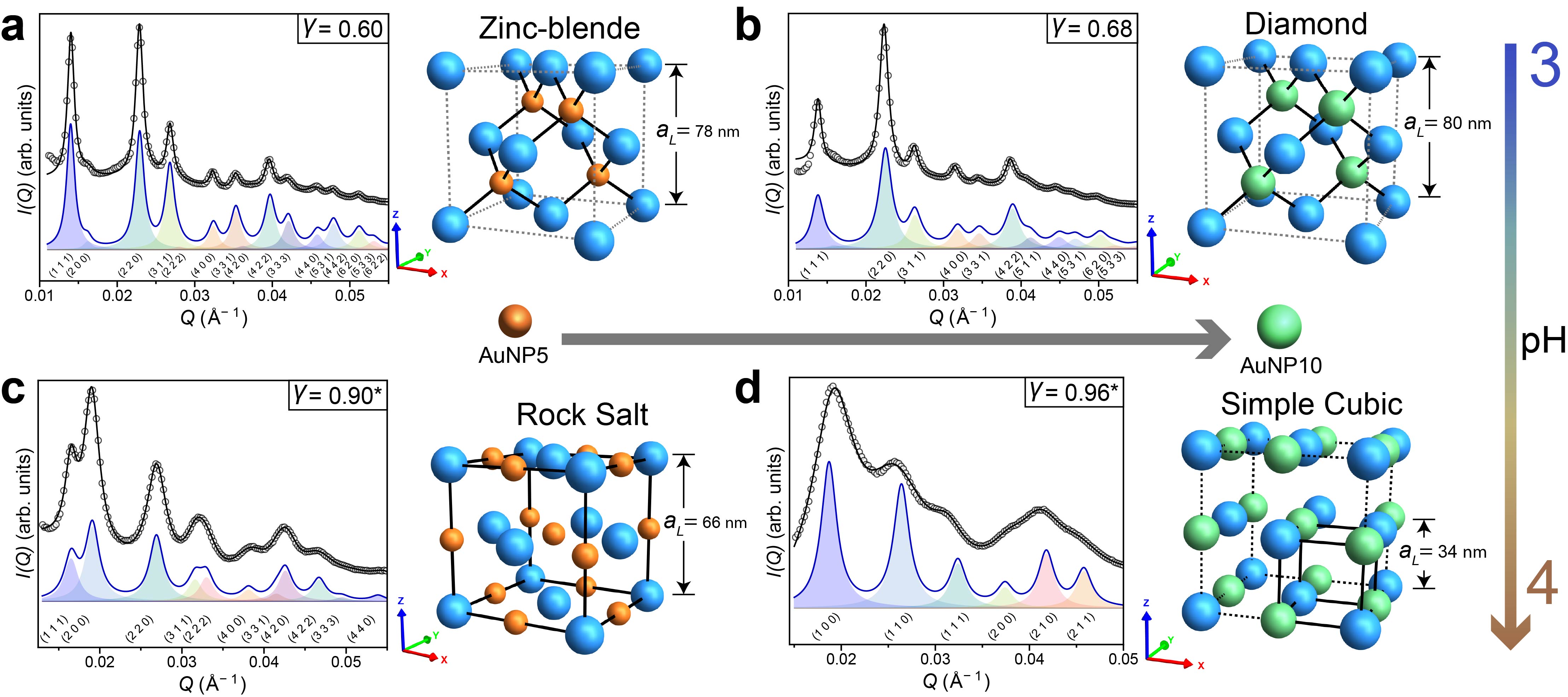}
   \caption{\textbf{Transition from Zinc-Blende to Diamond Superstructure.} (a) SAXS diffraction pattern ($I(Q)$ vs. $Q$) for COOH-PEG5k-Au5 and \ch{NH2}-PEG10k-Au10 at pH 3 (4:1 mixing ratio), fitted using a ZnS-like structural model. The solid line overlaid on the experimental data points (open circles) represents the fitted structure factor corresponding to each lattice. The modeled intensity profile (blue solid line) is shown below, with individual structure-factor contributions indicated by shaded colors. The corresponding Miller indices for each shaded contribution are provided below the plot.
   This ZnS-like structure can also be formed from the same binary system with mixing ratios of 1:1, 2:1, and 4:1, as shown in Fig. S7. (b) SAXS diffraction pattern for a 2:1 mixture of COOH-PEG5k-Au10 and \ch{NH2}-PEG10k-Au10 at pH 3, fitted to a diamond-like structural model. The only change from A to B is that the 5 nm NPs are replaced with 10 nm. This diamond structure can also form with mixing ratios of 1:1 and 4:1 for the same system, as shown in Fig. S9. (c) SAXS diffraction pattern for a 2:1 mixture of COOH-PEG5k-Au5 and \ch{NH2}-PEG10k-Au10 at pH 4, fitted to a NaCl-like structural model. This NaCl-like structure can also form with mixing ratios of 1:1 and 4:1 for the same system, as shown in Fig. S8. (d) SAXS diffraction pattern for a 1:1 mixture of COOH-PEG5k-Au10 and \ch{NH2}-PEG10k-Au10 at pH 4, fitted to a simple cubic structural model (blue solid line). This simple cubic structure arises from the identical core sizes (10 nm) of both grafted NPs, differentiating it from the binary NaCl-like structure. Illustrations of the ZnS-like, NaCl-like, diamond-like, and simple cubic superstructures are provided adjacent to their respective diffraction patterns, with the lattice constant ($a_{L}$) of each structure indicated in the illustrations. The SAXS intensity profiles, $I(Q)$, are normalized individually for each dataset and displayed on a linear scale. The $\gamma$ value for each mixture is indicated in the plot. The symbol $\ast$ represents the $\gamma$ value estimated from the nearest neighbor (NN) distance. 
    }
    \vspace{-0.3 cm}
\label{fig:Nacl_Sc} 
 \end{figure*}

Our approach begins with a binary system engineered to favor ZnS-like or NaCl-like superstructures. We use COOH-PEG5k-Au5 and \ch{NH2}-PEG10k-Au10 NPs, which yield a size ratio of \(\gamma \simeq 0.6\) optimal for a ZnS or NaCl-like superlattices assembly (see Fig.~\ref{fig:schematics}d). In addition, we define a parameter \(\theta\) as

\begin{equation}
\theta = \frac{{\rm MW}_{\rm COOH}}{{\rm MW}_{\ch{NH2}}},
\label{eq:theta}
\end{equation}

representing the ratio of molecular weights (MWs) of the grafted PEGs with different end groups. This parameter is valuable for constructing the experimental phase diagram of the assembled structures described below.

Fig. \ref{fig:Nacl_Sc}a shows the small-angle X-ray scattering (SAXS) diffraction pattern ($I(Q)$ vs. $Q$) for a 4:1 mixture of COOH-PEG5k-Au5 and \ch{NH2}-PEG10k-Au10 at pH 3. The diffraction pattern corresponds to a ZnS-like (zinc blende) structure, with the modeled intensity profile reproducing the experimental data. The ZnS-like structure can also form at different mixing ratios (1:1, 2:1, and 4:1), as shown in Fig. S7. Although ZnS-like assemblies consistently exhibit 1:1 stoichiometry, they can arise from mixtures with varying initial ratios of NPs, with any excess particles remaining unincorporated in solution. This behavior suggests that the assembly kinetics are not dependent on the initial particle ratios, as the system relaxes toward its thermodynamically favored stoichiometry, leaving excess particles unincorporated.
A more detailed analysis, presented in the SI Fig. S25, shows that the experimental pattern is best reproduced when site-exchanged defects are included in the ZnS framework, consistent with electrostatic stabilization of the open lattice. However, larger gold nanoparticles (AuNPs) (20 and 10 nm) (shown in Fig. S20,S21) yield good agreement between experimental diffraction patterns and calculated ZnS structure factors, demonstrating that increased particle size enhances structural stability and promotes the formation of well-ordered superlattices.

To achieve a reliable and well-defined diffraction pattern, we initially employed core NPs with distinct sizes (5~nm and 10~nm), thereby leveraging the electron density (ED) contrast between the two NP types to stabilize a ZnS-like lattice. For a binary system consisting of identical 10~nm core NPs (\(\gamma = 0.68\)), the lack of ED contrast transforms the ZnS-like into a diamond-like superlattice. Our work focuses on the gold NP lattice, treating the polymer framework, whose low ED is virtually invisible in X‑ray scattering, as negligible. In photonic applications, the NPs define the optical response, whereas the grafted polymers furnish the functional sites that organize and stabilize the active components. \cite{ho1990existence}
This applies to the SAXS diffraction pattern shown in Fig.~\ref{fig:Nacl_Sc}b, which is well described by a perfect diamond structure with minor deviations. In the ideal diamond lattice, four tetrahedral interstitial sites per unit cell are available. Our detailed fits indicate that, in some cases, these interstitials are partially occupied. As discussed in the SI Fig. S26, partial interstitial occupancy within the diamond lattice ($\sim 0.7$ NP per unit cell) provides an improved fit to the experimental diffraction pattern, suggesting minor deviations from ideal stoichiometry for enhanced structural stability.
These combinations indicate that optimal ZnS or diamond assembly occurs at \(\theta \simeq 0.5\). To further validate the robustness of our self-assembly strategy, we extended our investigation to binary mixtures of 20 and 10 nm AuNPs, functionalized with the same \ch{-COOH} and \ch{-NH2} terminated PEG ligands. By maintaining similar $\gamma$ and $\theta$ ratios, these larger particles also assemble into a well-ordered zinc-blende superlattice, as confirmed by the SAXS patterns in Fig. S20 and S21.

The stability of the zinc-blende structure offers a pathway for assembling diamond lattices. As shown in Fig. \ref{fig:schematics}a, the effective NP hard sphere diameter is defined as the sum of its core and the PEG corona thickness \cite{Zha2020}. By keeping the cores the same and tuning the PEG chain length to achieve \(\gamma\) values within \(\gamma_c^{\rm ZnS} \leq \gamma < \gamma_c^{\rm NaCl}\), diamond superlattices are formed. In contrast, when \(\gamma\) falls within \(\gamma_c^{\rm NaCl} \leq \gamma < \gamma_c^{\rm CsCl}\), a simple cubic structure is obtained.

The phase diagram in Fig.~\ref{fig:schematics}d depends on the parameter

\begin{equation}
\alpha = \frac{\Delta q_A}{q_A}\,\frac{\Delta q_B}{q_B},
\label{eq:alpha}
\end{equation}

where $\Delta q_A$ and $\Delta q_B$ denote the amount of surface charge specified within a defined solid angle, where the charge needs to be considered explicitly to account for correlations (See Fig.~\ref{fig:schematics}a and a more detailed derivation in SI). These localized charge correlations introduce a short‐range electrostatic correction to the hard‐sphere model, shifting the critical $\gamma$ values and thereby enhancing the relative stability of ZnS and NaCl lattices over CsCl. The dashed lines in Fig.~\ref{fig:schematics}d delineate a confined (\(\alpha,\gamma\)) window corresponding to the optimal conditions for the assembly of the above cubic superstructures. This highlights that tuning the NP surface charge (and hence \(\alpha\)) provides control over the resulting lattice symmetry. 

A further essential knob for controlling the superstructure assembly is the effective surface charge of the NPs, which can be finely tuned by adjusting the pH. As shown in Fig.~\ref{fig:Nacl_Sc}c, increasing the pH from 3 to 4 transforms the zinc-blende into a rock-salt superstructure. Similarly, the diamond-like superstructure observed at pH 3 (Fig.~\ref{fig:Nacl_Sc}b) transitions to a simple cubic configuration at pH 4 (Fig.~\ref{fig:Nacl_Sc}d). This simple cubic phase arises from NPs with identical 10~nm cores, where the loss of ED contrast diminishes the (1 1 1) peak characteristic of a rock-salt lattice. Furthermore, the SAXS pattern of the simple cubic phase exhibits a broad peak, indicative of mesoscale order and structural defects. Detailed definitions of these ordering regimes are provided in the Methods section. This imperfect, simple cubic lattice sets the stage for our subsequent pursuit of a defect-free simple cubic structure.

\begin{figure*}[!hbt]
 	\centering 
 	\includegraphics[width=1\linewidth]{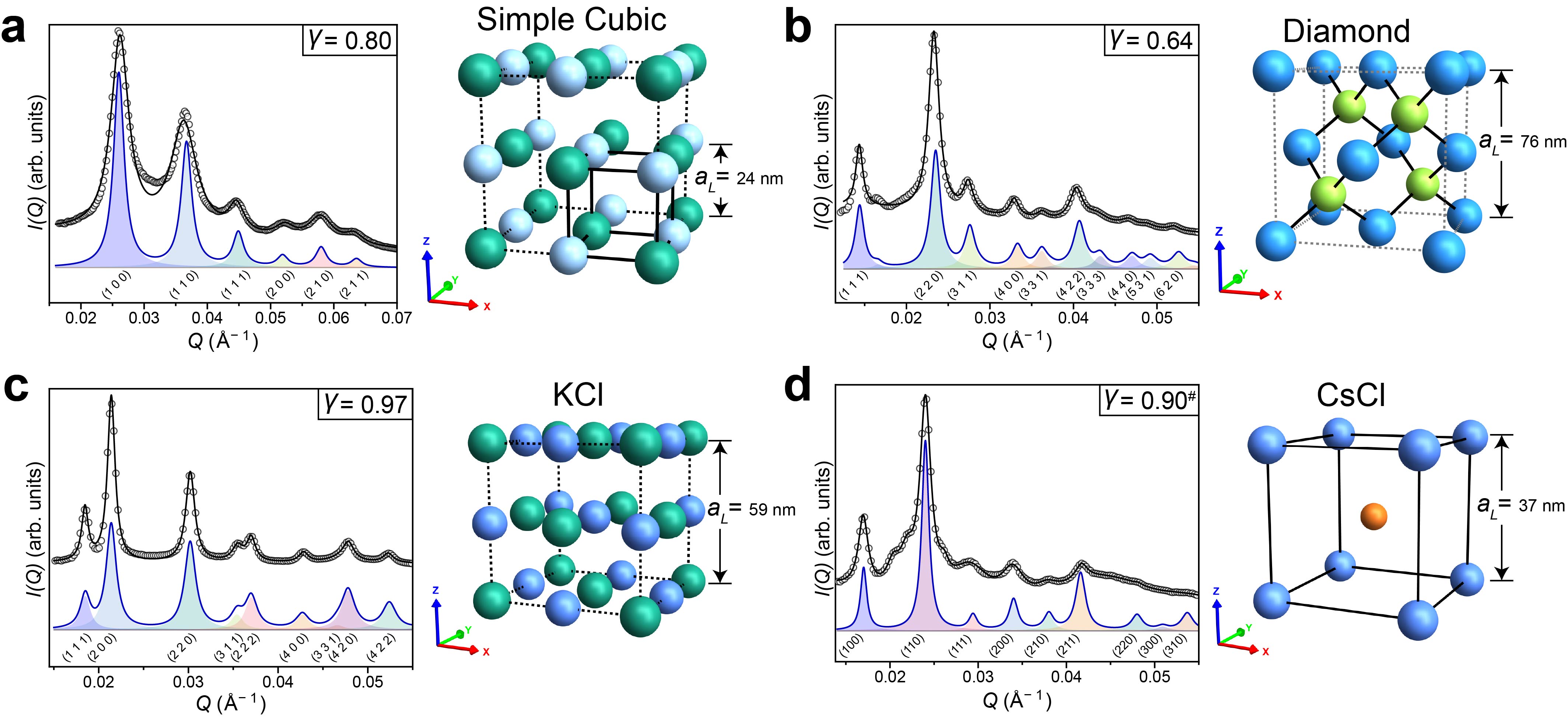}
      \caption{\textbf{Simple Cubic and Cesium-chloride–like Superstructures.} 
    (a) SAXS diffraction pattern ($I(Q)$ vs. $Q$) for a 1:1 mixture of COOH-PEG10k-Au10 and \ch{NH2}-PEG2k-Au10 at pH 3, consistent with a simple cubic structural model. This simple cubic structure is effectively equivalent to the NaCl-like arrangement; however, due to the similar core sizes of the NPs, X-ray scattering detects it as a simple cubic structure.  
    (b) SAXS diffraction pattern for a 1:1 mixture of COOH-PEG2k-Au10 and \ch{NH2}-PEG10k-Au10 at pH 3, consistent with a diamond-like structural model. This diamond structure is achieved by swapping the grafted PEG terminal groups in panel (a). The observed fluctuations in $\gamma$ arise from changes in the $D_{\rm H}$ of the \ch{-NH2}-terminated NPs.
    (c) SAXS diffraction pattern for a 1:2 mixture of COOH-PEG10k-Au10 and \ch{NH2}-PEG5k-Au10 at pH 3, consistent with a KCl-like structural model. Although a simple cubic arrangement derived from the NaCl-like structure might be expected, NP fractionalization favors the KCl-like structure.  
    (d) SAXS diffraction pattern for a 1:2 mixture of \ch{NH2}-PEG5k-Au10 and COOH-PEG5k-Au5 at pH 3, consistent with a CsCl-like structural model.  Experimental data are shown as open circles; structure-factor-Lorentzian fits are drawn as solid black lines through the data; and modeled intensity profiles from structural models are plotted below as blue solid lines with individual peak contributions shaded in color. The corresponding Miller indices for each shaded contribution are provided below the plot.
    Illustrations of the NaCl-like, diamond-like, CsCl-like, and simple cubic superstructures are provided adjacent to their respective diffraction patterns, with the $a_{\rm L}$ indicated in each illustration. The SAXS intensity profiles, $I(Q)$, are normalized individually for each dataset and displayed on a linear scale. The $\gamma$ value for each mixture is given in the plots, and the symbol $\#$ denotes the $\gamma$ value estimated from $D_{\rm H}$ as described in the SI.}

      \vspace{-0.3 cm}
\label{fig:SC_NaCl_Diamond} 
 \end{figure*}

Thus, we explore alternative parameter combinations by maintaining a constant \(\gamma\) while varying \(\theta\) (MWs of grafted PEG). Some combinations yield simple cubic short-range ordering (SRO).  However, setting \(\theta \simeq 5\) and \(\gamma \simeq 0.8\) (a condition that typically favors a rock-salt structure), the binary mixture of COOH-PEG10k-Au10 and \ch{NH2}-PEG2k-Au10 produced a simple cubic analog. Fig.~\ref{fig:SC_NaCl_Diamond}a shows the SAXS diffraction pattern of this assembly, which is accurately fitted to a simple cubic model.

By interchanging the terminal groups- swapping COOH and NH\(_2\), we stabilize a structurally well-defined diamond lattice. Fig.~\ref{fig:SC_NaCl_Diamond}b presents the SAXS diffraction pattern for a 1:1 mixture of COOH-PEG2k-Au10 and NH\(_2\)-PEG10k-Au10 at pH 3 (\(\theta \simeq 0.2\)), which is well-fitted to a diamond-like structure. This modification alters both the total and local charge on the NPs, as grafting density (and thus surface charge) depends on PEG chain length~\cite{kim2021effectchain,nayak2025effect}. Moreover, the effective size ratio \(\gamma\) shifts from 0.80 (yielding a simple cubic lattice) to 0.64 (to a diamond lattice), in accordance with the calculated phase diagram in Fig.~\ref{fig:schematics}D. These findings emphasize that maintaining low \(\theta\) and \(\gamma\) values is essential for stabilizing a diamond lattice, whereas higher values typically favor a rock-salt structure. We have successfully assembled a variety of diamond and zinc-blende lattices using these parameter regimes (see SI Figs. S13 and S15).

\begin{figure*}[!hbt]
 	\centering 
 	\includegraphics[width=1\linewidth]{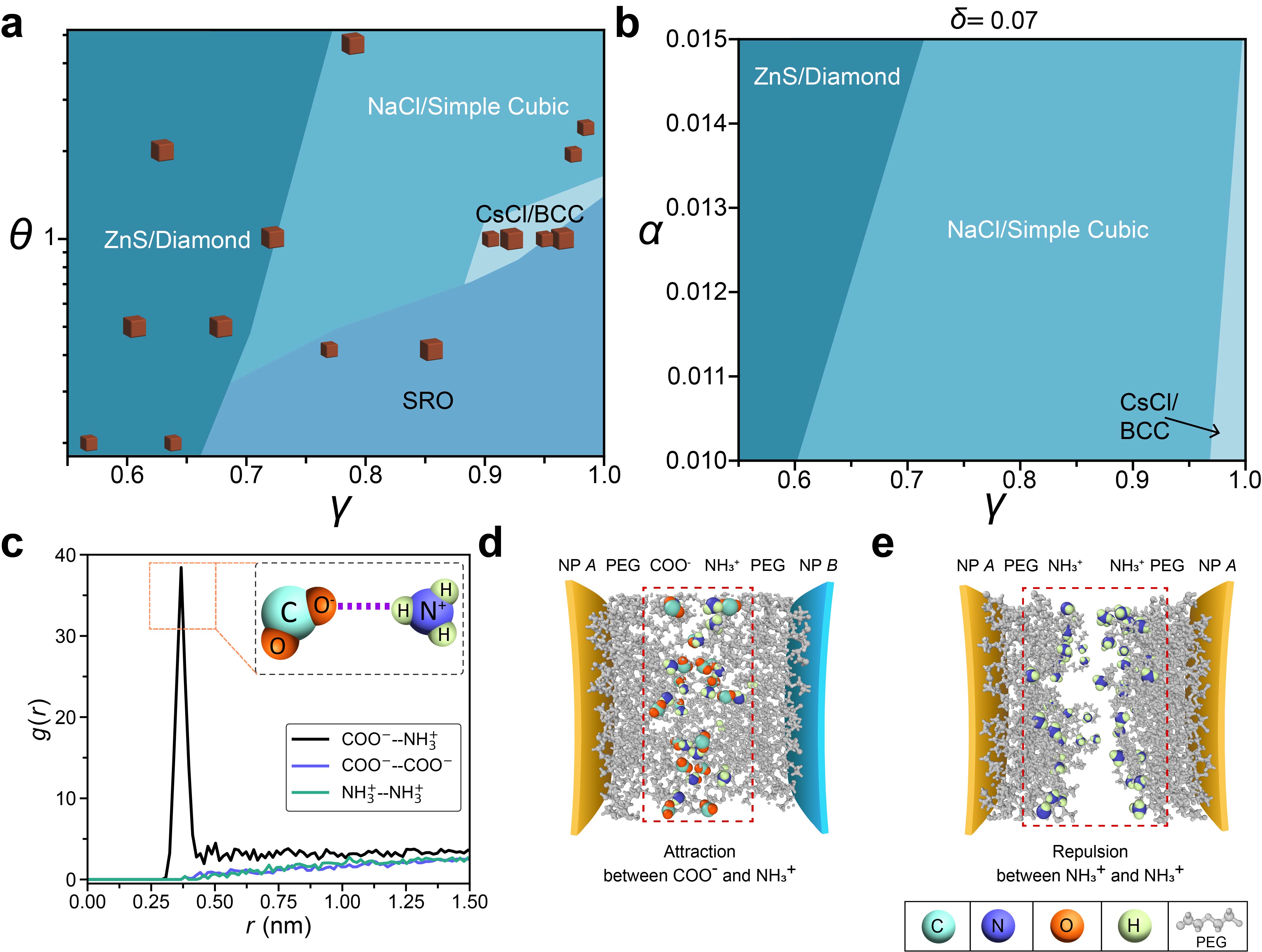}
      \caption{\textbf{Experimental, Theoretical Phase Diagrams and Molecular Simulations.} (a) SAXS-derived experimental phase diagram of cubic superstructures plotted as a function of \(\theta\) versus \(\gamma\). Larger cube symbols indicate binary mixtures containing 10~nm core NPs, whereas smaller symbols represent mixtures incorporating both 10~nm and 5~nm cores. (b) Calculated phase diagram of \(\alpha\) versus \(\gamma\) at a fixed \(\delta = 0.07\), which shows agreement with the experimental data. (c) Pair distribution function from simulations of PEG-functionalized flat walls with charged end groups illustrates electrostatic correlations. The prominent peak in the black curve reflects the hydrogen-bond-mediated attraction between COO\(^{-}\) and NH\(_3^{+}\) groups (inset: typical hydrogen bonding geometry). The blue and green curves lack peaks, indicating strong repulsion between like-charged groups. (d) Simulation snapshot highlighting the attractive interaction between NH\(_3^{+}\) and COO\(^{-}\) groups. (e) Simulation snapshot showing the repulsive interaction among NH\(_3^{+}\) groups.}
    \vspace{-0.3 cm}
\label{fig:phase} 
 \end{figure*}
 
Our next binary system exhibits the phenomenon of particle fractionalization, a process in which NPs are selectively sorted by size from their inherent polydispersity (See SI for detailed discussions). \cite{Lai2019,cabane2016hiding} Fig.~\ref{fig:SC_NaCl_Diamond}c shows the SAXS diffraction pattern for a binary mixture of COOH-PEG10k-Au10 and \ch{NH2}-PEG5k-Au10 with \(\theta \simeq 2\) and \(\gamma \simeq  0.97\), conditions that would typically yield a simple cubic pattern given the use of identical core NPs. However, the appearance of a weak (111) peak, characteristic of a NaCl-like structure (see Fig.~\ref{fig:Nacl_Sc}b), indicates that the assembly process is selective, assigning distinct particle sizes to different lattice sites through fractionalization. This slight size disparity results in a weak (111) reflection analogous to the differences observed between the diffraction patterns of NaCl and KCl, where reduced ED contrast in KCl produces a similar effect despite both crystallizing in an FCC structure. We surmise that the particle fractionalization is likely due to the redistribution of NPs across multiple assemblies when the system lies near a phase boundary.

Based on the phase diagram in Fig.~\ref{fig:schematics}d, we identified a narrow range of parameters that yield a CsCl-like structure. By selecting a binary mixture of COOH-PEG5k-Au10 and NH\(_2\)-PEG5k-Au5 NPs (\(\gamma \simeq 0.9\) and \(\theta \simeq 1\)), parameters that lie near the predicted coexistence region for NaCl and CsCl superstructures, and by tuning the pH to 3, we successfully assembled a CsCl-like superlattice. Fig.~\ref{fig:SC_NaCl_Diamond}d presents the SAXS diffraction pattern for this mixture at pH 3, which fits well to a CsCl-like structure. Small satellite peaks near the prominent (110) reflection indicate systematic defects or lattice elongation as discussed in detail in the SI. In contrast, using NPs with identical core sizes (10 nm and 10 nm) under the same grafting conditions to achieve a BCC superstructure resulted only in low-quality, short-range BCC-like ordering, as shown in S18. The phase diagram in Fig.~\ref{fig:phase}b justifies this finding by revealing that the CsCl/BCC phase is stable only within a very narrow parameter window. For completeness, the SI provides the detailed lattice parameters, nearest neighbor (NN) distances in Figs. S7-S19 and full width at half maximum (FWHM) in Table S8.  FWHM values were used to estimate the sizes of the crystalline domains, which are on the order of 1~µm or larger for the highest quality crystals, while lower quality domains are typically in the submicrometer range.

To rationalize and extend our experimental findings, we introduce the following semi‑quantitative free energy expression, which incorporates electrostatic correlations:

\begin{equation}\label{Eq:energy_model_rep}
E_L = \frac{{\cal N}_{P}}{2}\frac{e^2 q_A^2}{4\pi \varepsilon_0 D_A} \left( \frac{2 M_L}{1+\gamma}  + \frac{\alpha \, n_{AA}/2}{\delta + c_L(\gamma - \gamma_c)} \right).
\end{equation}

\noindent Here, \(\gamma\), \(\gamma_c\), \(\alpha\), and \(\delta\) are defined in Fig.~\ref{fig:schematics}a; \(M_L\) represent the Madelung energy of the corresponding lattice; \(c_L\) is a constant, and \(n_{AA}\) denotes the number of NNs for an $A$ NP. The second term in parentheses accounts for the short-range repulsion between $B-B$ NPs as \(\gamma\) approaches \(\gamma_c\). Increasing \(\alpha\) enhances this $B-B$ repulsion, thereby improving the stability of the ZnS structure by effectively shifting \(\gamma_c\) to a higher value (see Fig.~\ref{fig:schematics}d). A detailed discussion of this model is provided in the SI.

Understanding the assembly process requires elucidating the molecular‐level mechanisms governing NP interactions, particularly the attraction of $A-B$ NPs and the repulsion of $B-B$ (and $A-A$) NPs described in Eq.~\ref{Eq:energy_model_rep}. To this end, we performed all‑atom molecular dynamics (MD) simulations, directly quantifying the electrostatic and hydrogen‑bonding correlations that drive these forces. Our model system consists of planar walls (hypothetically NP surfaces) functionalized with PEG having COO\(^{-}\) or NH\(_3^+\) end groups, neutralized by electrolytes. The justification for this model and further supporting results are provided in the SI. Fig.~\ref{fig:phase}c shows the pair distribution function, \(g(r)\), for COO\(^{-}\)–NH\(_3^+\) pairs (solid black line), which exhibits a sharp peak indicating that roughly one-third of these end groups are bound. Additionally, almost half of these bound groups form bonds, with binding strength roughly independent of electrolyte concentration. This binding is enhanced by the formation of COO\(^{-}\cdots\)NH\(_3^+\) hydrogen bonds. In contrast, Fig.~\ref{fig:phase}d shows that equally charged end groups exhibit strong repulsion over distances up to 1 nm, thereby validating the cutoff distance, \(r_c\), used in our effective model (see Fig.~\ref{fig:schematics}a). Snapshot images in Fig.~\ref{fig:phase}d further illustrate that similarly charged end groups remain well separated, effectively overcoming potential screening by electrolytes or charge neutralization by counterions. Together, these simulations illustrate the $B-B$ correlations, confirming the repulsive term assumed in Eq.~\ref{Eq:energy_model_rep}, thus correcting the hard sphere model and leading to the phase diagram Fig.~\ref {fig:phase}b.


In summary, by leveraging pH-regulated surface charge and controlling the conformation of grafted polymers, we establish a generalizable, valence-free strategy for assembling open-framework cubic superlattices from binary mixtures of isotropic NPs. In these systems, a single set of design rules, encoded in the NP size ratio \(\gamma\) and the grafted ligand molecular weight ratio \(\theta\), directs crystallization into lattices analogous to rock salt, CsCl, zinc blende, diamond, and even the rare simple cubic phase. SAXS measurements indicate well-defined structural order, and our free-energy calculations and MD simulations provide quantitative support for the observed trends. Importantly, this approach is universally applicable to any NP–ligand system, enabling control over the lattice constant by tuning the core diameter and ligand molecular weight to adjust \(\gamma\) and \(\theta\).  We demonstrated that our approach is scalable, as we have assembled a zinc-blende lattice with NPs of 5, 10, and 20 nm in diameter, with a lattice constant ranging from 70 to 80 nm. Our versatile strategy establishes a broadly applicable framework for realizing open architectures without directional bonding, offering a general route to designing functional NP superlattices for photonic to catalytic applications.

\vspace{-0.4 cm}

\section{Methods}
\vspace{-0.2 cm}
\subsection{Materials}
Unconjugated citrate-capped AuNPs of nominal core diameters of 5 and 10 nm were purchased from Ted Pella Inc., and their size distributions were verified independently by transmission electron microscopy (TEM) ($5.8 \pm 0.6$ and $9.1 \pm 1$, respectively, See Figs. S5-S6) and by SAXS.\cite{nayak2023tuning} PEG with an average molecular weight of 2, 5, and 10 kDa, modified with a thiol (\ch{HS}) group at one end and either a carboxyl (\ch{-COOH}) or amine (\ch{-NH2}) group at the other, was purchased from Creative PEGworks (NC, USA) and used without further purification. HCl was purchased from Fisher Scientific and used without further purification. Milli-Q water (resistivity 18.2 M$\Omega$·cm at 25 \textdegree C) was used in all experiments.

\vspace{-0.4 cm}
\subsection{PEG Grafting to AuNPs Surfaces}
AuNPs are functionalized with HS-PEG-COOH or HS-PEG-\ch{NH2} using a ligand exchange protocol.\cite{nayak2025effect,zhang2017macroscopic} PEG ligands were dissolved in water and thoroughly mixed to form a homogeneous suspension. This PEG solution, in molar excess, was added to AuNP suspensions at a 1:1500 AuNP-to-PEG chain ratio for 5 nm AuNPs and a 1:6000 ratio for 10 nm AuNPs. The mixture was rotated overnight at $\sim$35 RPM on a Roto-Shake Genie (Scientific Industries, NY, USA) under continuous mixing to facilitate ligand exchange. Unbound PEG was removed from the PEG-functionalized AuNPs by three rounds of centrifugation. For 5 nm AuNPs, centrifugation was performed at 21,000 ×g for 90 minutes per round, while for 10 nm AuNPs, it was conducted at 20,000 ×g for 75 minutes. After centrifugation, the resulting PEG-grafted AuNPs were resuspended in Milli-Q water to create a stock suspension. The final concentration of the PEG-AuNP suspensions was determined by measuring absorbance using ultraviolet-visible (UV-vis) spectroscopy (NanoDrop One Microvolume, Thermo Fisher Scientific). Concentrations were adjusted to approximately 20 nM for 10 nm AuNPs and 80 nM for 5 nm AuNPs, as described below. 

The molarity of AuNP suspensions is calculated using the concentration data provided by the supplier (TedPella Inc.), which specifies the number of particles per milliliter of solution. The following equation is employed for the calculation:

\vspace{-0.2 cm}
\begin{equation}\label{Eq:molarity}
    \text{Molarity (M)} = \frac{\rm Number~of~particles~per~ml}{N_{\rm A} \times 10^{-3} \rm ~liters}
\end{equation}
where $N_{\rm A}$ is Avogadro's number.

\noindent For instance, the molarity of 10 nm-sized bare AuNPs is determined from the particle concentration provided in the vendor's technical data sheet, which is 5.7 $ \times 10^{12}$ particles/ml. Using Eq. \ref{Eq:molarity}, the molarity is calculated to be 9.46 nM. After determining the molarity of the bare AuNPs, UV-vis spectrophotometry is utilized to determine their absorbance at a wavelength of 525 nm. Assuming that the grafting of PEG does not influence the absorbance and that no aggregation occurs, the absorbance profiles of the PEG-grafted AuNPs can be compared with those of the bare AuNPs to determine the concentration of the grafted AuNPs in suspension. PEG grafting densities, determined by thermogravimetric analysis (TGA), are provided in the SI and are reproducible, independent of terminal group charge.\cite{nayak2025effect,zhang2017macroscopic,kim2023two,kim2022binary}

In this study, PEG-AuNP refers to PEG-grafted AuNPs in general, while \textit{x}-PEG\textit{y}-AuNP\textit{z} specifies the detailed characteristics of the NPs. In this labeling scheme, \textit{z} represents the core diameter of the AuNPs in nanometers (e.g., 5 or 10 nm), \textit{y} refers to the molecular weight of the grafted PEG (2, 5 or 10 kDa), and \textit{x} indicates the terminal functional group of the PEG chain (\ch{-COOH} or \ch{-NH2}). For example, COOH-PEG5k-Au10 describes 10 nm AuNPs functionalized with 5 kDa PEG terminated with a carboxyl group (\ch{-COOH}). Similarly, \ch{NH2}-PEG10k-Au5 denotes 5 nm AuNPs grafted with 10 kDa PEG terminated with an amine group (\ch{-NH2}). The complete labeling scheme and detailed properties of all PEG-grafted NPs are provided in Table S1. The legend for each PEG-AuNPs is depicted as a hard sphere in Fig. \ref{fig:schematics}b.

\vspace{-0.65 cm}
\subsection{DLS and $\zeta-$potential Measurements}
The $D_{\text{H}}$ of grafted NPs was measured by dynamic light scattering (DLS) on a NanoZS90 with Zetasizer software (Malvern, UK). $\zeta-$potential measurements, conducted on the same instrument, were used to quantify the surface charge of PEG-grafted AuNPs. These data are summarized in Table S1. Successful PEG grafting with distinct terminal groups (\ch{-COOH} or \ch{-NH2}) was confirmed by observing opposite surface charges and increased hydrodynamic sizes. Further details on PEG grafting and characterization are available in a recent publication.\cite{nayak2023ionic} While $D_{\text{H}}$ and the DLS-derived size ratio $\gamma$ provide objective, experimentally measurable design parameters to guide NP assembly, they are not expected to exactly predict the resulting lattice constants, which are governed by collective packing effects, polymer deformation, and electrostatic interactions.The $D_{\text{H}}$ values show a systematic increase for \ch{-NH2}-terminated particles compared to \ch{-COOH}-terminated ones, particularly at higher PEG molecular weights (e.g., 10 kDa), reflecting the more hydrated and expanded corona of NH$_2$-terminated PEG relative to the more compact conformation of COOH-terminated PEG.

\vspace{-0.4 cm}
\subsection{Surface Charge Manipulation}
To adjust the surface and overall charge of the grafted AuNPs, HCl was added to the suspensions, achieving final concentrations of 0.1, 1, and 10 mM. Since the small sample volumes required for SAXS hinder direct pH measurement, HCl concentrations and corresponding estimated pH values are specified in parentheses. pH values were verified by bulk titration of representative samples, showing consistency within $\pm$0.2 pH units, corresponding to the final pH values calculated from the amount of HCl added to the solutions. For SAXS, a calculated volume of AuNP suspension was mixed in a glass scintillation vial at the desired mixing ratio. Mixing ratio are calculated in terms of NP number ratios. Stock HCl solutions were added to the AuNP mixture,  thoroughly mixed, and incubated for $\sim 20$ minutes. The mixture was then loaded into a quartz capillary ($\sim 1.5$ mm diameter) for SAXS measurements.

\begin{figure*}[!hbt]
 	\centering 
 	\includegraphics[width=0.95\linewidth]{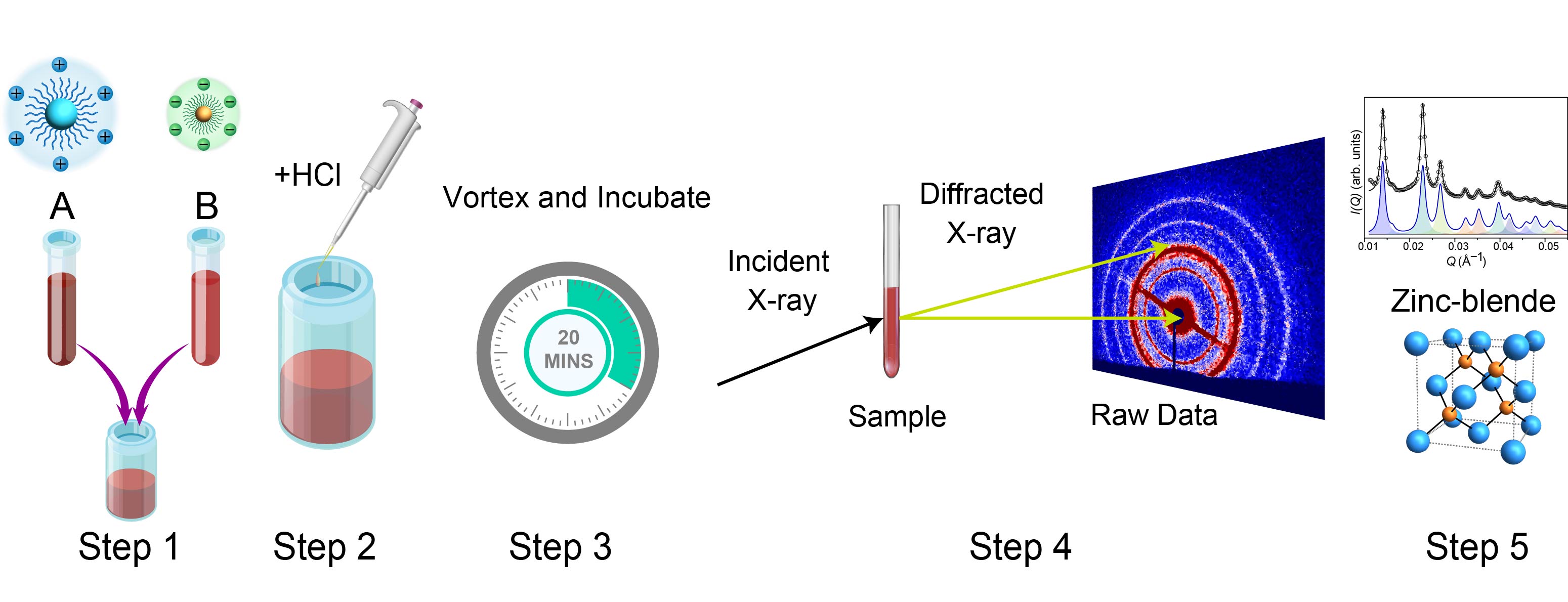}
    \caption{\textbf{Schematic illustration of the standard procedure for creating and characterizing superstructures by SAXS.} (1) Prepare the binary system using specified mixing ratios. (2) Adjust the pH of the mixture. (3) Thoroughly mix and incubate the suspension. (4) Load the mixture into a capillary and expose it to a monochromatic X-ray beam, resulting in a ring-like diffraction pattern characteristic of polycrystalline samples. (5) Convert the ring pattern to 1D \(I(Q)\) versus \(Q\) scattering plots and assign the corresponding superlattice structure.}

\label{fig:method} 
 \end{figure*}
 
\subsection{SAXS Measurements}
Fig. \ref{fig:method} illustrates the procedure for creating binary mixtures, conducting the SAXS measurements, and their structural analysis. 
\textit{In-situ} SAXS was employed to determine the structures formed in the suspensions at various pH levels and temperatures. Given the lower ED of the polymers compared to gold, the diffraction is dominated by the AuNPs. We employed AuNPs of two distinct sizes (10 nm and 5 nm) to achieve scattering contrast in our binary mixtures. In contrast, when the binary system comprises of NPs with identical core sizes (10 or 5 nm), the resulting superstructures exhibit a uniform, single-particle-like lattice (such as a diamond or simple cubic lattice).
For a general binary system, we designate the constituent particles as \(\mathcal{A}\) and \(\mathcal{B}\), where $D_A > D_B$, without specifying their precise sizes or the details of their PEG ligand terminations.

Synchrotron-based \textit{in-situ} SAXS experiments were conducted at beamline 12-ID-B of the Advanced Photon Source (APS), Argonne National Laboratory, and at the 11-BM CMS end station of the National Synchrotron Light Source II (NSLS-II), Brookhaven National Laboratory. The incident X-ray energies were 13.3 keV and 13.5 keV, respectively. Bench‑top \textit{in-situ} SAXS was performed using a liquid‑metal‑jet Ga alloy source with an incident X‑ray energy of 9.2 keV.

Incubated samples were transferred into $\sim$1.5 mm diameter quartz capillaries and positioned vertically in beamline-specific sample holders, normal to the incident beam. Data acquisition and processing were carried out following each beamline’s protocol. SAXS data were reduced to one-dimensional (1D) scattering profiles as a function of the wave vector magnitude \( Q \), where \( Q = 4\pi \sin \theta / \lambda \), with \( 2\theta \) being the scattering angle and \( \lambda \) the X-ray wavelength. Detailed experimental setup, data acquisition, and analysis protocols are available in previous studies.\cite{nayak2023assembling,kim2020temperaturenanorods,morozova2023colloidal}

The scattering profiles, \(I(Q)\), were background-corrected by subtracting signals from corresponding water samples, resulting in diffraction patterns resembling those of polycrystalline samples in suspension. The plotted scattering profiles are presented as raw \(I(Q)\) after only background subtraction; no form-factor correction \(P(Q)\) was applied owing to the dominance of long-range order scattering, and structural assignments are based on peak positions, with relative intensities noted but not used as the sole determinant. Data analysis followed standard procedures for X-ray diffraction of polycrystalline materials. For crystalline samples, peak positions were identified based on expected reflections for specific space groups, and structure factors were calculated and fitted to the data to determine the fundamental unit cell dimensions and particle occupancy.
Based on structural organization, we classified the assemblies into three regimes. Assemblies are defined as long-range ordered (LRO) when the experimental diffraction peaks are resolution-limited, typically corresponding to crystalline domains larger than 10 unit cells. SRO is assigned when the peak width is at least twice the instrumental resolution, corresponding to crystalline domains smaller than 5 unit cells. Meso-range order (MRO) describes intermediate cases, with crystalline domains of 5--10 unit cells~\cite{nayak2023ionic}. The FWHM values used for these classifications are summarized in Table S8.

\subsection{All-atom MD Simulations}

All-atom MD simulations using HOOMD-blue\cite{AndersonMe2008a} implemented through HOODLT\cite{Travesset2014, UpahTravesset2023}. HOODLT was used to generate the initial configurations, implement the force field, and analyze the results. The system consisted of two parallel plates composed of gold atoms, modeled as rigid bodies\cite{Nguyen2011} and grafted with PEG ligands terminating in charged end groups with the addition of salt (NaCl) in different concentrations, see justification of the model further below. The optimized potentials for liquid simulations (OPLS) force field\cite{Jorgensen1996}, implemented through Foyer\cite{Klein2019}, was used.
Three distinct systems were simulated: (a) a system with 98 PEG ligands in total 49 on each plate where one plate terminated in COO$^-$ and the other in NH$_3^+$ groups; (b) a system with 98 PEG ligands all terminating in COO$^-$ groups with Na$^+$ as counterions; and (c) a system with 98 PEG ligands all terminating in NH$_3^+$ groups with Cl$^-$ as counterions. Each system was simulated at varying plate separations ranging from 4.6 nm to 11.6 nm, and at salt molalities of 0.5 and 1.1 mol/kg, with SPC/E water used as the solvent. All simulations were first equilibrated in the NPT (constant pressure and temperature) ensemble for 3 ns using the MTTK thermostat and barostat to maintain constant temperature and pressure. This was followed by 40 nanoseconds of production run in the NVT (constant volume and temperature) ensemble with the same MTTK thermostat with a timestep of 1 fs. All simulations were conducted on the Bridges-2 high-performance computing cluster at the Pittsburgh Supercomputing Center.

\vspace{-0.5 cm}
\section{Data Availability}
All data supporting the findings of this study are available within the Article and its Supplementary Information. Source data for all main-text figures are provided with this paper as a Source Data file. Raw datasets, including SAXS measurements, electron microscopy images, DLS data, and simulation outputs, are deposited at the Harvard Dataverse under the accession DOI: 10.7910/DVN/JNGUK5. 

\vspace{-0.5 cm}
\normalem
\bibliography{Ref.bib,references.bib}

\vspace{-0.5 cm}
\section{Acknowledgements}
The authors thank Dr. Byeongdu Lee, Advanced Photon Source (APS), for helpful discussions. This work was supported by the U.S. Department of Energy (DOE), Office of Science, Basic Energy Sciences, Materials Science and Engineering Division. The research was performed at the Ames National Laboratory, which is operated for the U.S. DOE by Iowa State University under contract No. DE-AC02-07CH11358. The authors thank the 12-ID-B beamline staff team for the synchrotron beamline support at the APS, Argonne National Laboratory. Part of this research utilized beamline 12-ID-B of the APS, a U.S. DOE Office of Science user facility at Argonne National Laboratory, and is based on research supported by the U.S. DOE Office of Science, Basic Energy Sciences, under Contract No. DE-AC02-06CH11357. Part of this research used the Complex Materials Scattering (CMS) Beamline (CMS, Beamline 11-BM) of the National Synchrotron Light Source II, and the Materials Synthesis and Characterization Facility of the Center for Functional Nanomaterials (CFN), which is a U.S. DOE Office of Science User Facility, at Brookhaven National Laboratory under Contract No. DE-SC0012704.
\noindent Part of this work utilized the BRIDGES2 GPU cluster, allocated through MCB140071 from the Advanced Cyberinfrastructure Coordination Ecosystem: Services Support (ACCESS) program\cite{Boerner2023}, which is supported by National Science Foundation grants 2138259, 2138286, 2138307, 2137603, and 2138296. The work of ZM and SZ was supported by the U.S. DOE, Office of Science, Office of Basic Energy Sciences, Established Program to Stimulate Competitive Research (EPSCoR) Program under Award Number DE-SC0025585.

\vspace{-0.5 cm}
\section{Author contributions}
DV, AT, WW, and SM conceived and \newpage 
\noindent supervised the project. BN, DV, and WW designed and conducted the experiments and analyzed the data.  AT developed the theoretical and computational models, and PK performed all-atom molecular dynamics simulations. HZ and DN assisted in conducting and processing SAXS measurements. ZM and SZ carried out complementary electron microscopy imaging. BN, WW, AT, and DV wrote the manuscript. SM, DV, AT, and WW acquired project funding. All authors discussed the results and reviewed the manuscript.

\vspace{-0.5 cm}
\section{Competing interests}
The authors declare that they have no competing interests.

\end{document}